# A new light at the end of the tunnel: fiber gas discharge lasers

A. L. Love,[1] S. A. Bateman,[1] W. Belardi,[1][‡] F. Yu,[1][§] J.C. Knight,[1] D.W. Coutts,[2] C.E. Webb,[3] W.J. Wadsworth[1]*

[1]*Centre for Photonics and Photonic Materials, Department of Physics, University of Bath, BA2 7AY, UK*
[2]*MQ Photonics Research Centre, School of Mathematical and Physical Sciences, Macquarie University, NSW 2109, Australia*
[3]*Clarendon Laboratory, University of Oxford, OX1 3PU, UK*
[‡]*now with: Università degli Studi di Parma, Italy*
[§]*now with: Key Laboratory of Materials for High Power Laser, Shanghai Institute of Optics and Fine Mechanics, Chinese Academy of Sciences, Shanghai 201800, China*

*Corresponding author W.J.Wadsworth@bath.ac.uk*

**Optical fibers have emerged as a transformative platform for building better and more robust solid state lasers. However, the wavelengths available to these lasers are limited. Using hollow core optical fibers allows us to add gases as new potential gain media for fiber lasers, and also liberates the gas laser from the limits normally imposed by diffraction. To demonstrate the new technology, we present a fiber laser at 3.5 µm wavelength, using an antiresonant guiding hollow core optical fiber containing neutral xenon atoms pumped by an afterglow discharge of a helium-xenon mixture within a fiber of over 1 m in length. Laser action is confirmed through observation of polarization dependence, mode pulling and mode beating. Our results unlock a new breed of flexible fiber lasers operating at a plethora of wavelengths, many previous unavailable.**

Laser machining, attosecond science, optical communications, cold atom interferometers, etc. are all benefiting from the translation of bulk lasers into fiber form, using rare-earth doped optical fibers. Continuous wave and pulsed fiber lasers have reached beyond 1 kW average power[1, 2] and have become robust commercial systems used in industrial processing. Their efficiency, excellent beam quality and stability allow further power increases using coherent combination of several lasers to reach multi-kW average power, whilst maintaining ultrashort pulse duration and high peak power[3,4]. Such an architecture is used for attosecond science at the European Extreme Light Infrastructure facility[5]. Stable and low-noise optical fiber amplifiers are the backbone of long-haul optical communications and narrow linewidth fiber lasers are applied to cold atom interferometers[6]. Fiber laser wavelengths now extend into the mid-infrared [7,8] and ultrafast fiber lasers continue to develop [9,10] and are used in a growing number of research fields, such as ultra-precision spectroscopy using frequency combs [11,12].

Optical fiber lasers are becoming so successful because they are compact and robust; control of the laser alignment and laser beam quality do not rely on mechanical fixtures and bulk optical components. The key for optical fiber lasers is how light is guided in the core of the flexible optical fiber. Unfortunately conventional optical fibers are solid, which means that the benefits of the optical fiber geometry have not been available for an entire and very important class of laser – gas lasers.

Recent progress in hollow core optical fibers removes this restriction and significant milestones have been achieved. For instance, demonstrating coherent emission by gas molecules inside the fiber, gas-filled hollow optical fibers were applied to nonlinear wavelength conversion[13], including vacuum ultraviolet light generation[14]. Similarly, optically pumped gas lasers have also been demonstrated in hollow optical fibers[15,16,17]. Unfortunately, optical pumping requires narrow linewidth lasers as a pump source, which restricts the operation to only a small number of molecular gases, where appropriate pump lasers are available to match the narrow gas absorption lines. By contrast, gas lasers pumped by an electrical discharge offer a wealth of opportunities but they have never yet been realized in a fiber form.

Here we present the first demonstration of a fiber gas discharge laser, bringing together a large diameter hollow core optical fiber and a pulsed discharge in xenon. We show that discharge pumped gas lasers can be an effective fiber laser gain medium by demonstrating laser oscillation on a number of neutral xenon lines. The laser fiber is flexible and breaks the diffraction limit of a free-space gas laser by a factor of 500. Lasing is confirmed by observation of the output polarization and mode pulling. We chose xenon for this prototype gas fiber laser due to its low breakdown potential and the existence of a number of high gain laser transitions between the 5d and 6p neutral xenon energy levels. Our results open up the way for integrating a myriad of discharge pumped gas lasers into flexible optical fiber platforms. Due to the huge variety of gas laser spectral lines our technology will lead to an explosion of novel fiber lasers that will cover regions of the electromagnetic spectrum that have so far been unavailable.

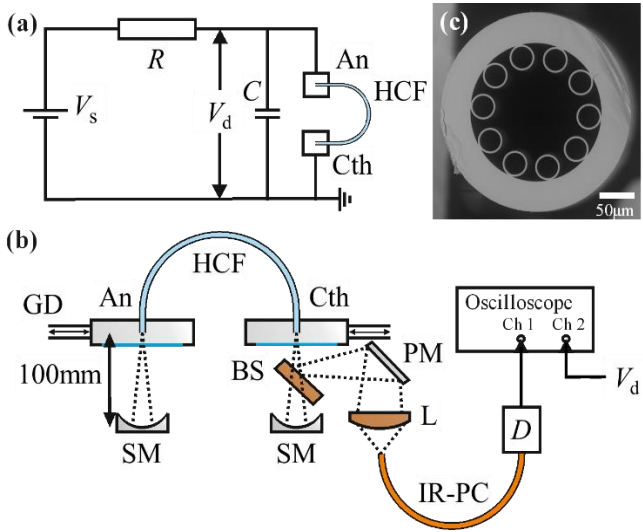

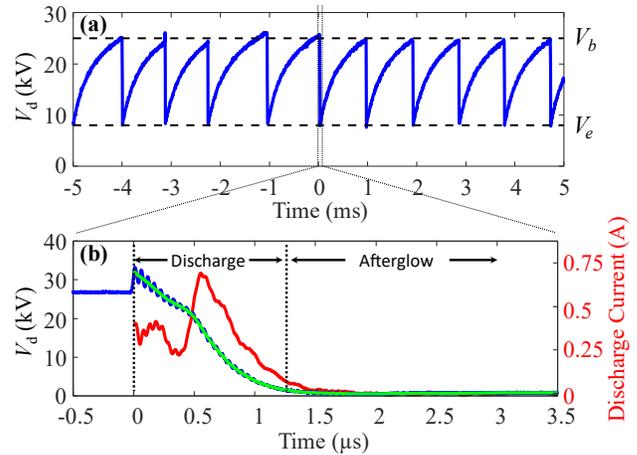

**Fig. 1. Experimental setup.** (a) Electrical excitation circuit: $V_s$, DC supply 0 - 40 kV; $R$, 60 MΩ ballast resistor; $C$, Stray capacitance ~10 - 20 pF; $V_d$, discharge potential; An, anode; Cth, cathode; HCF, hollow core fiber. (b) Optical layout of the laser cavity: SM, silver coated concave spherical mirror; BS, output coupling beam splitter; PM, planar mirror; L, lens; IR-PC, IR fiber patch cord; D, IR detector; GD, gas delivery (shown in further detail in Fig. S1). The custom made gas cells that house the electrodes feature $CaF_2$ (An) and sapphire (Cth) windows to provide optical access to the fiber ends. (c) Scanning electron micrograph of the cross-section of the hollow core fiber used. Overall diameter 273 μm, core diameter between inner circles 120 μm.

## Experimental

The xenon laser is an example of the family of neutral noble gas lasers, which includes neon in the common red helium-neon (HeNe) laser[18,19]. The energy levels of the noble gases all follow similar patterns. In an electrical discharge the population of the upper laser levels and the depopulation of the long-lived lower laser levels both depend on collisions between the atoms and excited electrons[20]. These collisions are enhanced at higher gas pressure. The stability of the glow discharges that provide the excitation is however dependent on the average electron energy (or electron temperature), which is in turn dependent on the pressure-diameter product $pd$[21]. The increase in pressure to maintain this product gives an improvement in laser gain with smaller bore diameters[22,23,24], which provides a further incentive to move gas lasers into a hollow optical fiber. To exploit the benefits of a small core to the full we must overcome diffraction, which places a fundamental limit on the useful length of a laser before the small beam spreads out to meet the walls of the tube. Some attempts were made in the early 1970's[25,26] to make use of grazing incidence reflection from the inside of a sub-mm inner diameter, $d$, glass tube in order to increase the useful length. However, such waveguides still suffer from a high loss that scales with wavelength, $\lambda$, as $\lambda^2/d^3$ [22,27] and have a stringent requirement for straightness, so this architecture gave only marginal benefits and could not provide the basis for a truly compact laser[22,26].

The first real opportunity for fiber gas lasers with no fundamental optical limit to their length was presented by the hollow core fiber in

**Fig. 2. Electrical characteristics of the discharge.** (a) Time trace of the discharge potential, $V_d$, over a 10 ms window as measured by a Tenma 72-3040 probe. The sawtooth-like voltage profile strongly resembles that of a Pearson-Anson oscillator. The black dashed lines indicate the breakdown and extinction voltages $V_b$ and $V_e$ respectively. (b) Time trace of the voltage across the discharge over a 4 μs window averaged over 128 pulses. The blue traces are the voltage as measured by the Tenma 72-3040 probe that was designed for DC purposes, so the fine features of the measurement such as the ringing in (b) are suspected to be an artefact. The green trace is the voltage waveform for $t > 0$ after smoothing to remove this ringing. The red trace is an estimate of the current, calculated by differentiating the smoothed voltage waveform.

1999 which used a photonic bandgap method of guidance[28], however the ~10 μm diameter cores proved too small to contain an electrical discharge. The recent development of low-loss, large core, hollow fibers based on antiresonant guidance[29,30,31] provides a long sought after solution to this problem, allowing long fiber lasers to be constructed with a discharge pumped gas as the gain medium.

The overall layout of the laser is shown in Fig. 1 (a) & (b). The fiber used for these experiments was an antiresonant design based on [31] It had a structure shown in Fig. 1 (c), with a core diameter of 120 μm. It was designed to guide in the region of the mid-infrared laser transitions of neutral xenon, particularly the high-gain line at 3.508 μm. The measured loss between 3.1 and 3.7 μm was below 0.2 dB/m, which is orders of magnitude lower than either the loss of the silica glass at this wavelength (~200 dB/m[32]) or a hollow capillary waveguide of these dimensions (~100 dB/m[27]). The bend loss was also low at less than 0.2 dB/turn for an 8 cm radius bend, and together these remarkable fiber properties enable the laser operation.

The hollow optical fiber was terminated in two gas cells which could be evacuated and filled with the laser gas mixture. An initial 12 mbar pressure of helium and xenon at a mixture ratio of 5:1 was chosen based on previous results[26], with scaling to the smaller diameter of the fiber core. One of the gas cells was brass and attached to electrical ground to form the cathode, the other cell was of insulating ceramic with a pin tungsten anode inserted. Both cells had infrared transmitting windows to allow the laser cavity to be formed with external concave mirrors. Extraction of the laser output was achieved by inserting a 50/50 beamsplitter, angled at 45° to the optical axis, between the cathode output and its respective mirror.

By applying a DC high voltage through a large ballast resistor (Fig. 1(a)), the negative differential resistance of the discharge together with the stray capacitance of the anode set up a self-oscillation similar to that of a Pearson-Anson oscillator[33]. The waveform of the discharge potential $V_d$ (Fig. 1 (a)) over a series of oscillations can be seen in Fig. 2 (a), where the oscillation repetition rate is about 1 kHz. At the start of one oscillation the stray capacitance of the anode is charged through the ballast resistor causing the anode potential, $V_d$, to increase until it reaches the discharge breakdown voltage, $V_b$, where the gas begins to ionize. The resulting drop in resistance causes the capacitor to discharge though the gas, producing a current pulse. The anode potential rapidly drops until it reaches the extinction voltage $V_e$, where the gas discharge extinguishes and the process repeats. Fig. 2 (b) is a zoomed in plot of one such discharge pulse. The current pulse has been calculated from the derivative of the voltage curve, using a capacitance of 14 pF calculated from the pulsing of Fig. 2 (a)(equations (S1), (S2)). As the voltage has been measured by a probe designed for DC applications, it is not known how accurate the shape of the voltage and current curves are. Nonetheless, estimates of the peak current and current pulse duration of the order of 1 A and 1 μs respectively provide sufficient understanding of the electrical pulse dynamics for the purposes of this study.

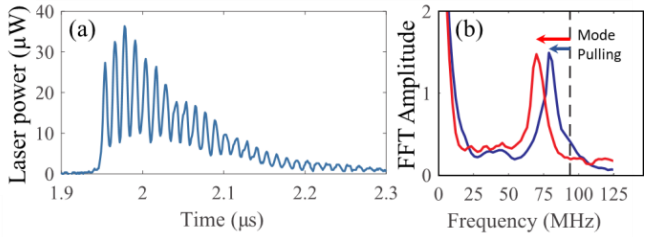

**Fig. 3. Demonstration of laser output and mode pulling.** (a) An example single laser pulse from a 1.6 m laser cavity. The optical pulse occurs in the afterglow of the discharge nearly 2 μs after the onset of the discharge (the time origin is set as in Fig. 2 (b)). The pulse exhibits clear oscillations from beating of two or more longitudinal modes of the laser cavity. These oscillations change in phase from pulse to pulse and are not visible when many pulses are averaged as in Figs 4 and 5. (b) Averaged magnitude from a set of 25 Fourier transforms of individual pulses from the same laser as (a) with good alignment (blue) and with a deliberately misaligned mirror at the anode (red). The measured beat frequencies are shifted from the empty cavity frequency of 94 MHz (dashed vertical line) through mode pulling.

The pulse duration and repetition rate were found to be dependent on the discharge length and the supply voltage respectively. While the time period between individual pulses was not perfectly regular, the discharges could be sustained over long periods of time. Sustained discharge operation beyond 30 minutes was only limited by impurities leaking into the system, not by the cataphoretic effects reported in previous DC excited helium-xenon systems[34].

The optical output was measured with a fast infrared HgCdTe photodiode and was found to occur in the afterglow of the current pulse, with a delay of up to 2 μs from the onset of the discharge (Fig. 3 (a)). This is a consequence of the high current density during the discharge. The output was confirmed to be predominantly at 3.5 μm wavelength by a monochromator (see Fig. S4). Oscillations within individual laser pulses were observed as a consequence of mode beating, as the cavity is long enough for more than one longitudinal cavity mode to fall within the gain bandwidth. This confirms that there is true laser oscillation in the cavity rather than single- or double-pass amplified spontaneous emission (ASE). As the phase of the beat signal varies from pulse to pulse the beat frequency was measured by taking the average of the magnitude in frequency space of the fast Fourier transform of 25 measured pulses. The observed beat frequency between the modes seen in Fig. 3 (b) is close to 80 MHz, which is substantially shifted from the expected spacing of modes in an empty cavity of 1.6 m total length ($\Delta\nu$ = 94 MHz). This mode pulling is a result of the narrow gain bandwidth and a relatively high saturated gain required in the cavity to overcome optical losses at the windows and output coupler[26,35].

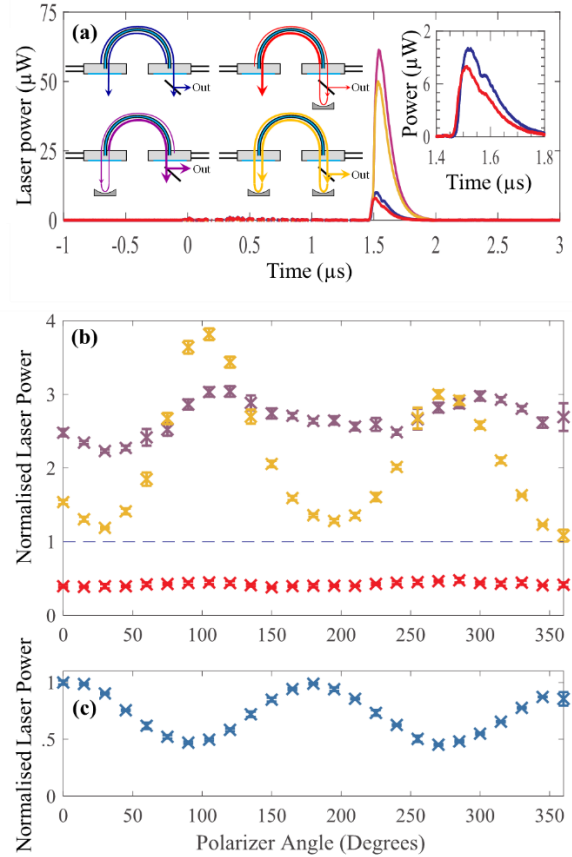

**Fig. 4. The laser output is polarized.** (a) Optical pulses of predominantly 3.5 μm radiation from discharges in 1 m fiber length at 12 mbar pressure and 10:1 He:Xe ratio. This pressure and gas ratio were chosen as they exhibit the highest output powers achieved. Traces are averaged over 128 pulses, so do not show mode beating. Different colors indicate different mirror configurations; blue – single-pass ASE, red – reverse double-pass ASE, purple – forward double-pass ASE and orange – complete laser cavity. Inset: zoom-in of the single-pass ASE, and reverse double-pass ASE pulses. (b), (c) Measurements taken from the output of a 1.5 m long fiber filled with a 5:1 mixture of helium and xenon at a pressure of 12 mbar by a CW InSb detector through a polarizer rotated in steps of 15° from the S-polarization at 0°. (b) Normalized forward double-pass (purple), reverse double-pass (red) and full cavity (orange) outputs powers. Each data point has been normalized to the corresponding single-pass ASE reading at that polarizer angle (represented here by the dashed blue line) to correct for the polarization dependence of the output coupler. A reduced output from the cavity is observed in the maximum at 270°, which is likely to be an ageing effect as the data was taken in order of increasing angle. (b) Single-pass ASE output signal normalized to the 0° data point. This shows the 2:1 polarization dependence of the reflection of the output coupler as the single pass ASE is expected to be unpolarized.

Fig. 3 (b) shows that the mode pulling increased when one of the cavity mirrors was partially misaligned, because an increased cavity loss requires a higher saturated gain to compensate. The gain bandwidth was calculated to be 265 ± 8 MHz from Doppler and pressure broadening[36,37], which is significantly lower than that of conventional rare-earth doped fiber lasers. The mode pulling allows us to calculate the saturated gain and hence the round-trip loss in the cavity. The inferred round-trip loss ranged from 14 ± 3 dB for the aligned cavity to 26 ± 3 dB in the misaligned case. Using the known optical properties of the cavity components (Section S2, S8) we calculate that approximately 40% of the available light was coupled back into the fiber at each end by the concave mirrors when well aligned.

Altering the arrangement of the system by sequentially blocking the cavity mirrors provided more insight into the gain and saturation characteristics of the laser (Fig. 4 (a)). With both curved mirrors blocked, there was substantial single-pass ASE as a result of the high gain of this emission line and the long gain medium. When the cathode concave mirror was aligned to the fiber, the resulting double-pass ASE was predominantly backward-going with respect to the output direction (hence forth referred to as a 'reverse double-pass'), and the reduced ASE observed from the first pass in the forward direction indicates that the gain becomes saturated. When the anode concave mirror was aligned instead, the resulting forward double-pass ASE was very strong. Aligning both mirrors completed the laser cavity.

Because the output beam is from the mode of the optical fiber and the gain is saturated on just a double-pass, little change in power and no change in beam profile is expected when the cavity is formed to make a laser instead of just ASE. Indeed, the laser power is seen to be slightly lower than the forward double-pass ASE output. The existence of a polarization dependence in the output coupler however means that there is a shift in the polarization state of the output once the laser cavity is completed, as the laser selects the lower-loss P-polarization. Evidence of this can be seen in Fig. 4 (b),(c), which plots normalized average output powers for the different cavity configurations for different polarizations. While the reverse and forward double-pass outputs exhibit little polarization dependence, for the completed laser cavity maximum output powers were observed in the P-polarizations, at polarizer angles of 90° and 270°.

The output power of the laser was optimized by varying both the total pressure and the ratio of the gases. The peak powers of the output from a discharge in a 1 m long fiber in the single-pass ASE and laser cavity arrangements are plotted against total pressure for various ratios of helium to xenon in Fig. 5 (a) and Fig. 5 (b) respectively. At each ratio the measurements were made at increasing pressures starting at the lowest pressure at which breakdown could be achieved, before a final repeat measurement at 12 mbar was made as a control to ensure that no degradation had occurred over time.

There is a clear reduction in output power with increased pressure, although this is more pronounced in the single-pass measurements where there was no external influence from the mirror alignment which showed some drift over the time of the measurements. There is also an increase of output power at lower xenon content, although at ratios of 10:1 and 20:1 the discharge would not start at pressures below 12 and 16 mbar respectively. Fig. 5 (c) shows the laser pulse output power traces for a 7:1 mixture at various pressures. In addition to the decreasing peak power, we also observed an increased time delay between the electrical and optical pulses with an increased pressure. Fig. 5 (d) is a similar plot but for various mixtures at a pressure of 12 mbar. The same power-pressure relationship on display suggests that it is in fact the partial pressure of xenon that plays the biggest role in the observed behavior. Measuring the pulse repetition frequency and pulse duration showed little change with pressure or gas composition, confirming that the observed variations of peak power will also be reflected in pulse energy and average power (see Fig. S6). The inverse correlation between power and pressure is likely to stem from the inverse relationship between electron-ion recombination rates and electron temperature[38]. Afterglow lasers are in part reliant on electron-ion recombination to populate the upper levels[21] and slower recombination would result in a weaker, more delayed pulse.

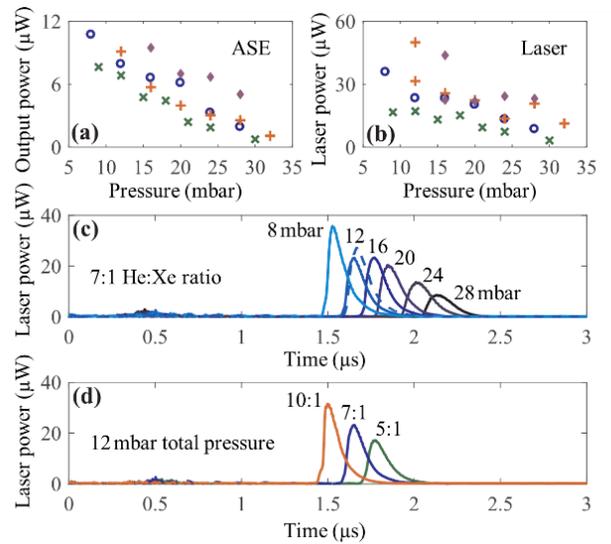

**Fig. 5. Optimization of the lasing process.** Pulse characteristics from a 1 m fiber filled with gas ratios of 5:1 (green cross), 7:1 (blue circle), 10:1 (orange plus) and 20:1 (purple diamond) He:Xe. Measured peak powers of the single-pass ASE (a) and laser cavity (b) outputs. The peak powers are observed to decrease with increasing pressure. (c) Laser output power traces for a 7:1 mixture of He:Xe at pressures of 8, 12, 16, 20, 24 and 28 mbar (light to dark; note the dotted repeat measurement at 12 mbar was a control, taken last). (d) Laser output power traces for a 12 mbar total pressure with different gas mixtures. The higher total pressures (c) and higher partial pressure of xenon (d) both show longer time delays between the electrical pulse (set at 0 s) and the light pulse, in addition to the lower peak powers (b).

Evidence of ASE and laser action was also observed for the 3.11 and 3.36 μm lines of xenon (see section S9). Discharges have also been achieved in neon-xenon, helium-neon, and argon gases within lengths of the same hollow core fiber. While discernible lasing signal has yet to be detected from helium-neon mixtures or argon, lasing was achieved in neon-xenon mixtures. With the lower ionization potential of neon compared to that of helium, the potential difference required to achieve breakdown is lower and so a smaller partial pressure of xenon could be used, to exploit the pressure dependence observed earlier. The resulting laser using a 200:1 neon to xenon ratio at a pressure of 10 mbar produced a similar magnitude of output power to the helium-xenon mixtures, but used discharges that were less than half the length (see Fig. S7).

## Conclusions

This initial laser demonstrates the ability of discharge pumped gas lasers to be operated inside flexible optical fibers, adding the properties of the enormous variety of gas laser lines to the range of available fiber lasers, potentially including many visible, IR and UV laser wavelengths.

The power demonstrated here is limited largely by the length of discharge achievable through longitudinal DC pumping. External transverse microwave excitation or combined microwave and DC excitation would be expected to offer discharges in a longer fiber, or multiple discharges along the fiber length. The operable length of the laser could then be increased until limited by fiber loss at lengths of the order of 100 m for the current state of the art hollow fibers, substantially increasing output power.

**Funding.** Engineering and Physical Sciences Research Council (EPSRC) (EP/I011315/1).

**Acknowledgment**. We thank Prof. T.A. Birks for useful discussions during the course of the project.

See Supplementary information for supporting content. All data underlying the results presented in this manuscript can be found at [39].

# A new light at the end of the tunnel: fiber gas discharge lasers: supplementary material

A. L. LOVE,[1] S. A. BATEMAN,[1] W. BELARDI,[1‡] F. YU,[1§] J.C. KNIGHT,[1] D.W. COUTTS,[2] C.E. WEBB,[3] W.J. WADSWORTH[1]*

[1]Centre for Photonics and Photonic Materials, Department of Physics, University of Bath, BA2 7AY, UK
[2]MQ Photonics Research Centre, School of Mathematical and Physical Sciences, Macquarie University, NSW 2109, Australia
[3]Clarendon Laboratory, University of Oxford, OX1 3PU, UK
‡now with: Università degli Studi di Parma, Italy
§now with: Key Laboratory of Materials for High Power Laser, Shanghai Institute of Optics and Fine Mechanics, Chinese Academy of Sciences, Shanghai 201800, China

*Corresponding author W.J.Wadsworth@bath.ac.uk

---

**This document provides supplementary information to "A new light at the end of the tunnel: fiber gas discharge lasers." Included are full details of the materials and methods used in the demonstration of the new laser and additional data showing the polarization dependence of the laser, evidence of the laser lines present and further experimentation with the gas mixture.**

## 1. Vacuum and gas handling system

The layout of the vacuum and gas handling system can be seen in Fig. S1. The system was evacuated to a pressure of $10^{-5}$ mbar by an Oerlikon Turbolab 80 vacuum pump, a two stage pump featuring a Turbovac SL80H turbo pump backed up by a dry diaphragm pump. The narrow fiber cannot be evacuated to this level, so a purge procedure was also used to remove any impurities within the fiber that would adversely affect the discharge. The gases were supplied by bottles of helium (N5.5 purity), neon (N5.0 purity) and xenon (N5.0 purity). The vacuum pressure was monitored by a penning gauge while the pressure in the fiber was monitored at both ends by capacitance gauges, which provide a gas-independent readings down to $10^{-2}$ mbar.

The ends of the fiber were sealed in specially designed gas cells. The cathode cell was made of brass, and featured a 2 mm thick uncoated Sapphire window while the anode cell was made out of Macor machinable ceramic and featured a 5 mm thick uncoated $CaF_2$ window. The rest of the system was constructed of aluminum KF fittings and stainless steel Swagelok fittings, with a plastic tube connecting the anode cell to isolate it electrically. Once filled, the valves around the fiber were closed so that the rest of the system could be evacuated before the voltage was turned on. This was to avoid any unwanted discharging happening within the system itself, particularly through the plastic tubing connected to the anode cell.

If the gas flow is assumed to be viscous and laminar, it would take approximately 10 s to fill 1.5 m of a 120 μm inner diameter fiber to 12 mbar[1]. However at these pressures the mean free path of the gas atoms is of the order of the fiber inner diameter, and so the flow is only laminar and fast when purging at higher pressures. At the pressures used for discharges a settling time longer than 10 s is required and so the system was left for several minutes. To combat the slow rate of free-molecular flow at low pressures and remove any impurities trapped in the fiber, the system was flushed with helium multiple times at pressures greater than 10 mBar before use.

## 2. Optical system

All the optical components were chosen for their transmission properties in the mid-IR spectral region. At 3.5 μm the reported transmission of uncoated Sapphire is 88%, and of $CaF_2$ is 95%, with surface reflections being the main source of loss. The output coupler was a 5 mm thick $CaF_2$ window with a broadband beamsplitter coating (ThorLabs BSW511), providing transmissions of 39% (61% reflectance) in the S-polarization, and 69% (31% reflectance) in the P-polarization at 3.5 μm at 45° incidence. The light was coupled back into the fiber using two silver coated spherical mirrors, each with a 100 mm radius of curvature and a reported reflectance of 97% at 3.5 μm (ThorLabs CM254-050-P01). The thick angled output coupler will induce some astigmatism that reduces the efficiency of coupling light back into the fiber at the cathode end.

The output light was collected by a 100 μm core multimode $InF_3$ fiber patch cable (ThorLabs MF12, 100 μm core diameter, ≤0.45 dBm$^{-1}$ attenuation from 2 to 4.6 μm) via a planar silver coated mirror and an anti-reflection coated $CaF_2$ lens (Thorlabs LA5183-E), and delivered to a VIGO PVI-3.4 IR photovoltaic detector. The detector had a response time of ≤2 ns, a spectral range of 2.9 to 4 μm and a responsivity of 2.5 A W$^{-1}$ and was connected to a VIGO SIP-100-250M-TO39-NG preamplifier with a gain of 10 kV A$^{-1}$. The detector output was displayed on an oscilloscope and averaged over 128 pulses, with the trigger set by the falling discharge voltage. The electrical pulses of the discharge caused a large amount of electrical noise to occur in the signal traces, though this was easily removed by subtracting a dark trace (with the detector blocked) from the active traces.

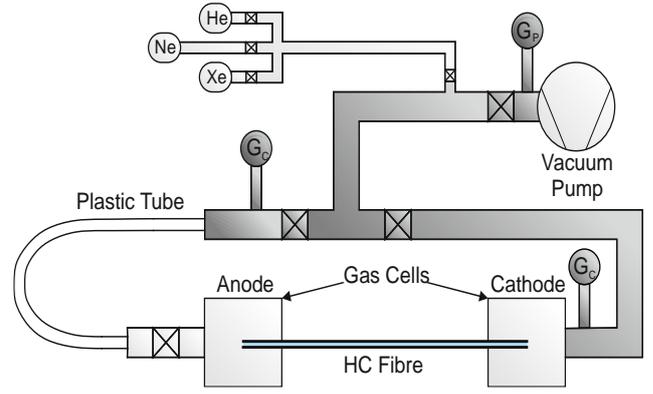

**Fig. S1.** Schematic diagram of the vacuum and gas filling system. Valves are indicated by crosses, and capacitance pressure gauges by Gc and the penning pressure gauge by Gp. The two gas cells are custom made. The anode, made of ceramic, is on the left is connected to the rest of the system via a stainless steel ¼ inch Swagelok valve and ¼ inch plastic tubing to isolate it electronically. The stray capacitance inferred from the electrical response is believed to come in part from this metal valve. The cathode, made of brass, is connected to the rest of the system by aluminum KF fittings (indicated by a darker shading). The gas bottles are connected via ¼ inch stainless steel tubing and Swagelok fittings.

## 3. Fiber specifications and loss measurements

The fibers used in the experiments are of the same design as those presented by Belardi *et al.* [2], with an outer diameter of 273 μm, a core diameter of 120 μm, resonator diameters of 35 μm and wall thicknesses of 2.8 μm.

The loss of the fibers was measured by using the cut-back method, where a transmission spectrum was taken from a broadband light source through a long length of fiber, before a length of fiber was cut off without disturbing the input coupling and another spectrum taken. By dividing the initial measurement by the subsequent shorter measurement, the loss over the removed length of fiber was calculated. The bend losses were measured in a similar way by first measuring a straight fiber transmission, before coiling the fiber into a loop of known diameter and re-measuring the transmission. The fiber used suffers from losses no greater than 0.4 dB m$^{-1}$ throughout its main guidance band between 3-4 μm, with measured transmission and bend losses of 0.16 dB m$^{-1}$ and less than 0.2 dB/turn at 3.5 μm.

## 4. Electrical system and capacitance calculation

The 40 kV required for the electric discharges were supplied by an UltraVolt 40A24-P30 power supply, capable of producing 30 W of power at up to 0.75 mA of current. The ballast resistance was provided by high voltage 1 MΩ resistors in series. With a 60 MΩ ballast resistor, the probe used to measure the discharge voltage must be of very high impedance in order not to disturb the electrical system. A Tenma 72-3040 probe provides the required high input impedance of 10 GΩ, but was designed for high voltage DC applications (up to 40 kV) so the fast pulses cause a ringing artifact in the measurements.

The stray capacitance $C$ was calculated to be around 10-20 pF from the time period of the pulses $T$, the ballast resistance $R$, the supply voltage $V_s$, the breakdown voltage $V_b$ and the extinction voltage $V_e$ from

$$T = RC \ln\left[\frac{V_s - V_e}{V_s - V_b}\right]. \tag{S1}$$

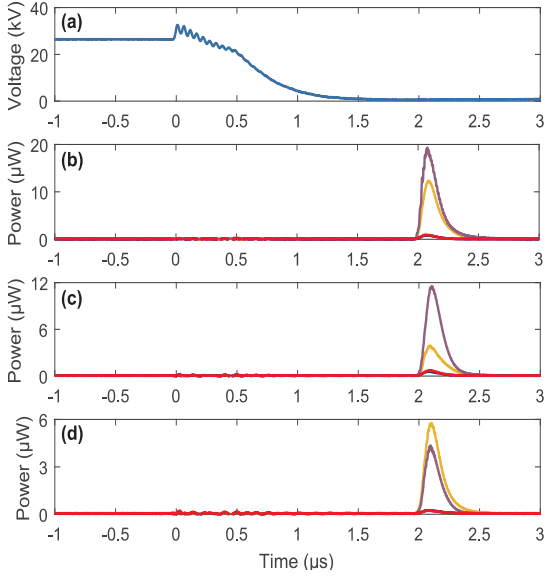

**Fig. S2.** Representative pulses for 1.4 m fiber length, 12 mbar pressure and 5:1 He:Xe ratio. Traces averaged over 128 pulses. (a) Discharge potential. (b), (c), (d) Optical pulses for unpolarized detection (b), and detection of S- (c) and P-polarization (d) respectively. Traces have not been corrected for the polarization dependence of the output coupler, and hence are plotted on different scales. Blue – single-pass ASE (calculated intra-cavity polarization ≈45% P-polarized), red – reverse double-pass ASE (≈39% P-polarized), purple – forward double-pass ASE (≈43% P-polarized) and orange – the complete laser cavity (≈74% P-polarized). The intra-cavity polarizations were calculated by dividing the P-polarization peak powers by the sum of the S- and P- peaks, all of which had been corrected for the output coupler polarization dependence.

The current pulse featured in Fig. 2 was calculated by numerically differentiating the voltage curve and inserting the values into

$$I = I_S - \frac{dV}{dt}C, \quad (S2)$$

where $I_S$ is the current flow from the power supply, measured to be around 0.25 mA.

## 5. Polarization

With the output coupler exhibiting a polarization dependence, the system will preferentially lase on the lower loss P-polarization provided that P-polarized light at the output coupler returns as P-polarized light after a round trip in the cavity. Fortunately this is the case here. In addition to possessing no deliberate birefringence, the lack of any material stress-induced refractive index perturbations in a hollow core fiber means that there is also no unintended birefringence. The geometry of the path of the fiber will however cause the polarization to rotate, but this rotation is reversed when light travels back along the same path [3,4].

The polarization of the laser was measured using a Thorlabs LPMIR050-MP2 polarizer. The S-, P- and unpolarized pulsed output can be seen in Fig S2.

## 6. Broadening mechanisms

The broadening of the gain bandwidth in a gas laser can be calculated as the sum of the Doppler broadening and the pressure broadening. The natural linewidth is generally much smaller than either of these. The Doppler broadening is well defined and calculated by [5]

$$\Delta f_{\text{Dop}} = \sqrt{\frac{8kT\ln(2)}{mc^2}}f_0, \quad (S3)$$

where $k$ is the Boltzmann constant, $T$ is the temperature of the xenon atoms, estimated to be around 300 ± 30 K, $m$ is the mass of a xenon atom, $c$ is the speed of light and $f_0$ the center frequency. This has been calculated to be 93 ± 5 MHz for the 3.51 µm xenon line. The pressure broadening in a 1:900 mixture of xenon and helium has been empirically shown to be

$$\Delta f_{\text{Pres}} = 4.7 + (18.6 \pm 0.7)p, \quad (S4)$$

for a total pressure $p$ in Torr [6]. For a 12 mbar pressure this yields a pressure broadening of 172 ± 6 MHz. While this is for a different ratio of the gases, the change is not expected to be significant.

## 7. Mode beating and Fourier analysis

With a significant contribution from Doppler broadening and a cavity length great enough so that multiple longitudinal modes fall within the gain bandwidth, the laser is likely to operate on several longitudinal modes. With each mode exhibiting a slightly different frequency, the modes will beat together creating some signal modulation in the output with a frequency equal to the spacing between the modes. This is calculated by taking the Fourier transform of the output.

To reduce the noise, averages were taken of the signal, however as the phase of the modulation on each shot is both varying and unknown, an average of the magnitude of the Fourier transforms from 25 single shot recordings was taken. All Fourier analysis was performed with the MATLAB package.

## 8. Mode pulling and gain calculation

In lasers that exhibit very high gain and narrow bandwidth, like the 3.51 µm xenon laser, a population inversion within the cavity induces a slight increase in the refractive index of the laser medium. This index increase results in laser cavity modes that are closer together in frequency [7]. This mode pulling frequency shift can be used to calculate the saturated gain of a homogeneously broadened laser as

$$\alpha = \frac{\pi \Delta \nu_W}{l}\left(\frac{1}{\Delta \nu'} - \frac{1}{\Delta \nu}\right), \quad (S5)$$

where $\Delta \nu_W$ is the gain bandwidth, $l$ the length of the gain medium, $\Delta \nu'$ the 'hot' cavity beat frequency and $\Delta \nu$ the empty cavity beat frequency. For the results shown in the text of $\Delta \nu_W = 265 \pm 8$ MHz, $\Delta \nu' = 80 \pm 2$ MHz, $\Delta \nu = 94 \pm 2$ MHz and $l = 1.4 \pm 0.1$ m, $\alpha = 1.1 \pm 0.2$ m$^{-1}$ giving a round trip amplification factor of $\exp(2\alpha l) = 20 \pm 10$, and a round trip cavity loss of $14 \pm 3$ dB. Combining the losses (section S2) of all the individual optical components together gives an ideal round trip cavity loss of 72.5%, meaning that each mirror is coupling back in approximately 40% of the light.

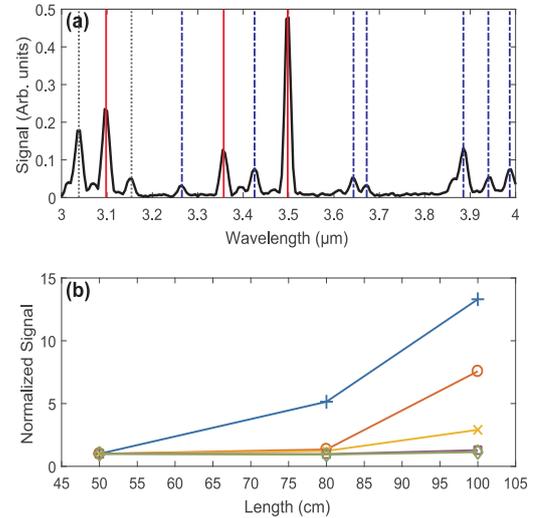

**Fig. S3.** (a) Spectra for single-pass discharges in a 5:1 ratio of helium and xenon at 12 mbar inside a 50 cm length of fiber at a 20 nm resolution. The gray dotted lines identify lines that have not been reported to lase, the blue dashed lines identify lines that have been reported to lase and the red lines identify the lines that have shown evidence of gain in this system (as well as previously). (b) Length dependence of signal for the 3.04 (green diamonds), 3.11 (orange crosses), 3.36 (red circles), 3.51 (blue plusses) and 3.89 µm (grey squares) lines. Each is normalized to the output signal for 0.5 m length.

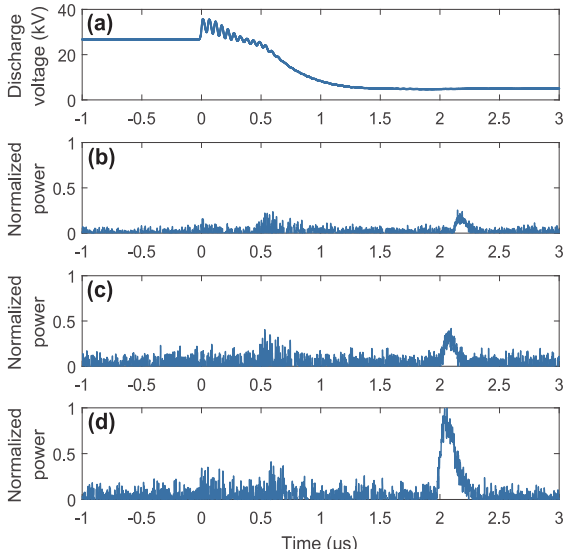
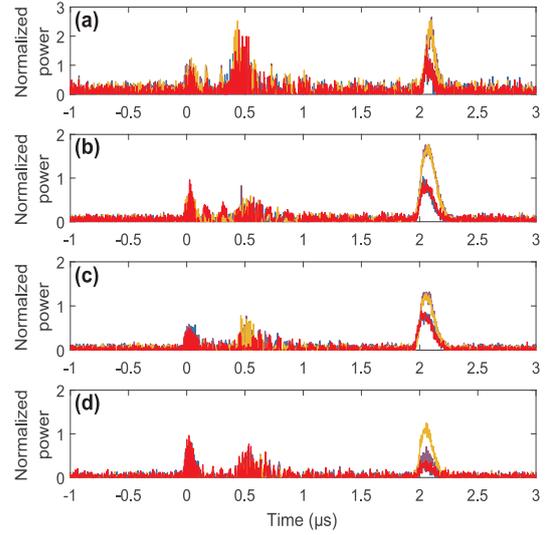

**Fig. S4**. Representative pulses for 70 cm fiber length, 12 mbar pressure and 5:1 He:Xe ratio. Traces averaged over 64 pulses. (a) Discharge potential. (b), (c), (d) Optical pulses for the single-pass output powers measured through the Bentham TMc300 monochromator at 3.11 μm (b), 3.36 μm (c) and 3.51 μm (d). The power traces have been normalized to the 3.51 μm line peak power. The 3.11 and 3.36 μm measurements were made with a 20 nm band pass, while the 3.51 μm measurement was made with a 10 nm band pass.

**Fig. S5**. Optical output for a 1.4 m fiber length, 12 mbar pressure and 5:1 He:Xe ratio. Traces averaged over 64 pulses. Different colors indicate different cavity configurations; blue – single-pass ASE, red – reverse double-pass, purple – forward double-pass and orange – complete laser cavity. (a) Unpolarized, (b) S-polarized and (c) P-polarized output powers at 3.36 μm respectively normalized to the unpolarized single-pass peak. (d) Unpolarized output power at 3.1 μm normalized to the single-pass peak.

## 9. Spectral analysis of the fiber ASE and laser output

In preliminary work [8] a CW InSb detector was used in combination with a monchromator (Bentham DTMc300 with a 300 grooves/mm grating) to take spectra of the output from single pass discharges between 3 and 4 μm for 0.5 m of fiber (see Fig. S3 (a)). These spectra revealed light to be present at a number of xenon lines. Further analysis of each individual line (see Fig. S3 (b)) revealed a super-linear growth of signal with length on the 3.11, 3.36 and 3.51 μm lines, a good indication of gain being present. In addition to this, the lack of this super-linear growth on the 3.04 and 3.89 μm lines suggested that the signal increase on the first three lines was not just a fluorescence effect, but amplified spontaneous emission.

Fig. S4 displays the single pass signals for the 3.11, 3.36 and 3.51 μm lines respectively using the fast IR detector through the monochromator, verifying that they are all present in the afterglow. The 3.51 μm line is clearly dominant, even though the manufacturer's quoted detector response is lower at 3.5 μm. Fig. S5 then shows the laser output for the 3.11 and 3.36 μm lines for the four arrangements of the cavity as in Fig. 4a for the 3.51 μm line. The measurement for the 3.36 μm line includes a polarization analysis and shows a strong response to the cavity mirrors although less evidence of saturation and polarization selection than was seen for the 3.51 μm line. The signal from the 3.11 μm line was too low to perform a similar experiment, but the observed response from blocking and revealing the cavity mirrors suggest that it too is a laser oscillator not just amplified spontaneous emission.

## 10. Additional pulse characteristics

Figure S6 displays additional data taken while the pressure and gas mixtures were varied which includes the pulse repetition rate and full width at half maximum (FWHM). The repetition rate was calculated by taking the Fourier transform of a 2 s window of the voltage waveform, whilst the FWHM was measured directly from the average light pulse waveform. They both show little correlation with pressure, suggesting that the variation of the laser's average power will closely follow that of its peak power.

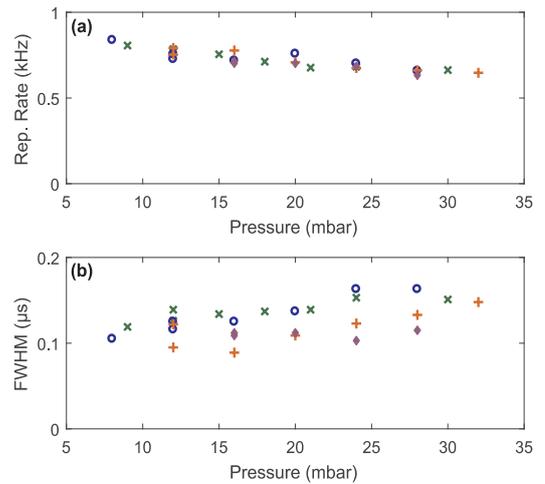

**Fig. S6**. (a) Pulse repetition rate and (b) light pulse duration (FWHM) from a 1 m fiber filled with gas ratios of 5:1 (green cross), 7:1 (blue circle), 10:1 (orange plus) and 20:1 (purple diamond) He:Xe. Both show minimal change, hence the average power will closely follow the peak power.

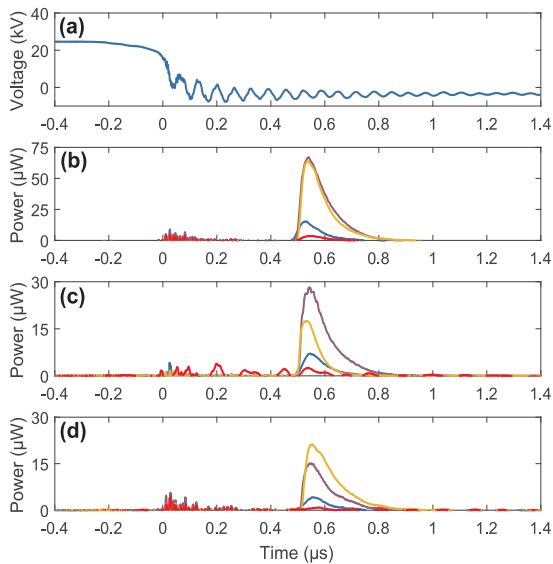

**Fig. S7.** Representative pulses for 65 cm fiber length, 10 mbar pressure and 200:1 Ne:Xe ratio. Traces averaged over 128 pulses. (a) Discharge potential. (b), (c), (d) Optical pulses for unpolarized detection (b), and detection of S- (c) and P-polarization (d) respectively. Traces have not been corrected for the polarization dependence of the output coupler, and hence are plotted on different scales. For the red trace in (c) the subtraction of electrical noise using a 'dark' trace was less effective than usual.

## 11. Experiments with different gases

Electrical discharges were also achieved within the same hollow core fibers in helium-neon, neon-xenon and argon gas mixtures. Replacing the helium-xenon with neon-xenon enabled the discharges to be performed with a lower partial pressure of xenon, and so as a result of the inverse pressure-output power relation demonstrated in Fig. 3 a higher power could be achieved, even in a fiber that was half the length.

Fig. S7 shows the output traces from a 200:1 mixture of neon-xenon at a pressure of 10 mbar in a 65 cm long piece of fiber. As before, the unpolarized, S-polarized and P-polarized outputs (Fig. S7 (b)-(d) respectively) are shown for each of the four cavity configurations and show preferential lasing in the P-polarization. A noticeable difference is the voltage waveform (Fig. S7 (a)), which could have been a result of the shorter length or the different gas mixture. The current pulse is much shorter than that seen in Fig. S4 which increases the measurement artifacts of the slow probe. The optical output still occurs in the afterglow region, but at a much smaller time delay after the shorter current pulse. An optical output signal was also detected at 2.026 µm, where a further xenon laser line exists, which showed a response from the mirrors.

The helium-neon discharges were achieved in a 5:1 mixture at a pressure of 42 mbar inside 55 cm of fiber. No signal was detected at the 3.39 µm laser line, possibly due to the differing excitation methods in helium-neon.

As the current pulses in helium-xenon were estimated to be of the order of 1 A (which gives a current density of over 10000 A cm$^{-2}$), argon was tested to look for evidence of the emission lines of ions. Using a 5 mbar pressure of argon inside 50 cm of fiber, electrical breakdown was achieved and a number of the notable argon ion laser lines between 454 and 528 nm were measured. No response from either mirror was observed, although this was not entirely unexpected with the high fiber attenuation in this spectral region.